\documentclass[manuscript]{emulateapj}
\usepackage{natbib}
\def\msol{\,{\rm M}_\odot}              

\def\ga{\,\hbox{\hbox{$ > $}\kern -0.8em \lower 1.0ex\hbox{$\sim$}}\,}
\def\la{\,\hbox{\hbox{$ < $}\kern -0.8em \lower 1.0ex\hbox{$\sim$}}\,}

\begin{document}

\title{
Analytical star formation rate from gravoturbulent fragmentation}

\author
{Patrick Hennebelle }
\affil{Laboratoire de radioastronomie, UMR CNRS 8112,\\
 \'Ecole normale sup\'erieure et Observatoire de Paris,\\
24 rue Lhomond, 75231 Paris cedex 05, France }

\and

\author{Gilles Chabrier}
\affil{\'Ecole normale sup\'erieure de Lyon,
CRAL, UMR CNRS 5574, 69364 Lyon Cedex 07,  France\\
School of Physics, University of Exeter, Exeter, UK EX4 4QL}

\date{}

\begin{abstract}
We present an analytical determination of the star formation rate (SFR) in molecular clouds, based on a time-dependent 
extension of
our analytical theory of the stellar initial mass function (IMF). The theory yields SFR's in good agreement with 
observations, suggesting that turbulence {\it is} the dominant, initial process responsible for 
star formation. In contrast to previous SFR theories,  the present one does not invoke an ad-hoc 
density threshold for star formation; instead, the SFR {\it continuously} increases with gas density, naturally 
yielding two different characteristic regimes, thus two different slopes in the SFR vs gas density relationship, 
in agreement with observational determinations.
Besides the complete SFR derivation, we also provide a simplified expression,  which reproduces reasonably well 
the complete calculations and can easily be used for quick determinations of SFR's in cloud environments.
 A key property at the heart of both our complete and simplified theory is that the SFR  involves a 
{\it density-dependent dynamical time}, characteristic of each collapsing (prestellar) overdense region in the cloud,
 instead of one single mean or critical freefall timescale.
Unfortunately, the SFR also depends on some ill-determined parameters, such as the core-to-star mass 
conversion efficiency and the crossing timescale. Although we provide estimates for these parameters, 
their uncertainty hampers a precise quantitative determination of the SFR, within less than a factor of a few.
 \end{abstract}


\section{Introduction}

The determination of the star formation rate (SFR) in molecular clouds and in galaxies
is one of the main challenges  of star formation theory. 
In the modern paradigm of star formation, stars form out of prestellar cores which result from the 
gravo-turbulent fragmentation
of molecular clouds (e.g. MacLow \& Klessen 2004).
Within the past few years, two analytical approaches have emerged, aiming at characterizing the SFR 
issued from the probability density function (PDF) of density fluctuations induced by turbulence in the 
cloud (Krumholz \& McKee 2005-KM, Padoan \& Nordlund 2011-PN).
Both theories rely on (i)  a density threshold, whose nature differs 
in each case, for star formation, (ii) {\it one} characteristic dynamical timescale, defined either at the  cloud's mean density or at
 the threshold density.
In this Letter, we derive a SFR, based on our IMF analytical theory (Hennebelle \& Chabrier 2008-HC08, 2009-HC09,
 Chabrier \& Hennebelle 2011), and show that (i) this theory yields SFR values in good agreement with 
observations, 
(ii) there is no {\it a priori} density threshold for star formation; instead, the SFR {\it continuously} 
increases with gas density, with indeed two different regimes.
 We also show that the exact value of the SFR depends on the combination of some ill-determined parameters, notably
 the core-to-star efficiency and the crossing timescale, whose uncertainties, and dependence 
upon cloud conditions,  hamper an exact determination of the SFR.

\section{Star formation rate: theories}
\label{sfr_sec}

We first summarize the previous SFR theories by  KM and PN. We then briefly present the SFR derived
from a time-dependent extension of our theory of star formation, which will be  presented in details in a forthcoming paper. Finally we present a simplified version of this theory which, alternatively, 
can be seen as an improved KM or PN theory. 

Following Krumholz \& McKee (2005), 
we define the dimensionless 
{\it star formation rate per free-fall time}, $SFR_{ff}$, as 
the fraction of cloud mass converted into stars
 per cloud {\it mean} free-fall time, $\tau_{ff}^0$, i.e.:
$SFR_{ff} = {\dot{M}_*\over M_c} \, \tau_{ff}^0$,
where $\dot{M}_*$ denotes the {\it total star formation rate} arising from a cloud
of mass $M_c$, size $L_c$  and  mean density $\rho_0$.

\subsection{The Krumholz and McKee theory}
\label{krum}
According to various simulations  of hydrodynamic or MHD supersonic turbulence, the density PDF is well 
represented  in both cases by a lognormal form,
\begin{eqnarray}
\label{Pr0}
{\cal P}(\delta) &=& {1 \over \sqrt{2 \pi \sigma_0^2}} 
\exp\left(- { (\delta - \bar{\delta})^2 \over 2 \sigma_0 ^2} \right) , \;
 \delta = \ln (\rho/ \rho_0 ) \\
 \bar{\delta}&=&-\sigma_0^2/2 \;
,  \; \sigma_0^2=\ln (1 + b^2 {\cal M}^2),
\nonumber
\end{eqnarray}
where ${\cal M}$ is the Mach number and $b \simeq 0.5$
(Federrath et al. 2010). 

The essence of the KM analysis is to assume that there is a critical 
density, $\rho_{crit}$, above which star formation is occuring. Then, the SFR  (eqn.(20) of KM)
  is simply obtained by estimating the fraction of 
gas with density larger than $\rho_{crit}$,
\begin{eqnarray}
SFR_{ff} = \epsilon {\tau_{ff}^0 \over \tau_{ff,cr} \phi_t} 
\int ^\infty _{\ln \widetilde{\rho}_{crit}}
\widetilde{\rho} {\cal P}({\delta}) d\widetilde{\delta},
\label{sfr_simple}
\end{eqnarray}
with  ${\widetilde \rho} = \rho / \rho_0$. KM further assume that $\tau_{ff,cr} \simeq \tau_{ff}^0$.

In this expression,
$\epsilon$ is the (supposedly mass-independent) efficiency with which the mass within 
the collapsing prestellar cores is converted into stars. 
Calculations (e.g. Matzner \& McKee 2000, 
Ciardi \& Hennebelle 2010) as well as observations (e.g. Andr\'e et al. 2010) suggest that $\epsilon \simeq 0.3-0.5$. The parameter $\phi_t$ corresponds to the time needed for a self-gravitating 
fluctuation to be replenished. KM estimate it to be of the order of a few, in agreement with the 
analysis we propose in the appendix.

In KM, $\rho_{crit}$ is determined from the condition that
 the corresponding 
Jeans length must be  equal to the sonic length.  Their underlying
assumption is that turbulent support will be too efficient to 
enable star formation at scales larger than the sonic length. This yields
$\widetilde{\rho}_{crit, KM} = (\phi_x \lambda_{J0}/\lambda_s)^2$, where
$\phi_x$ is a coefficient of order unity, $\lambda_{J0}$ is the Jeans length
at the mean cloud density and $\lambda_s$ is the sonic length.

\subsection{The Padoan and Nordlund theory}
The expression obtained by  PN is similar 
to the KM one, stated by eqn.~(\ref{sfr_simple}), except that they 
do not assume $\tau_{ff}^0 = \tau_{ff,cr}$, but instead
$ \tau_{ff,cr}/ \tau_{ff}^0 = \sqrt{\tilde{\rho}_{crit}}$,
as  indeed comes out from the integral in eqn.~(\ref{sfr_simple}). They consider that both $\epsilon$ and $\phi_t$
are equal to 1, except in the magnetized case where 
they argue that $\epsilon \simeq 0.5$ (which appears to be the main reason for the reduced SFR
in the magnetized case). With eqns.~(\ref{Pr0}) and (\ref{sfr_simple}),
this yields (eqn.(30) of PN) 
\begin{eqnarray}
SFR_{ff} = {\epsilon \over 2 \phi_t} \widetilde{\rho}_{crit}^{1/2}
\left[ 1 + {\rm erf} \left( 
{ \sigma_0^2 -2 \ln (\widetilde{\rho}_{crit}) \over 2^{3/2} \sigma_0} \right) \right].
\label{sfr_simple2}
\end{eqnarray}

The main difference with the KM model, however, resides
in the choice of $\rho_{crit}$. In PN,
this latter is obtained by requiring that the corresponding 
Jeans length be equal to the typical thickness of the shocked layer, inferred 
by combining isothermal shock jump conditions and a turbulent
velocity scaling $v \propto l^{0.5}$. This yields 
$\widetilde{\rho}_{crit, PN} \simeq 0.067\, \theta^{-2} \alpha_{vir} {\cal M}^2$,
where $\theta\approx 0.35$ is the ratio of the cloud size over the turbulent integral scale and
$\alpha_{vir}$ is the virial parameter,  
$\alpha_{vir}=2E_{\rm kin}/E_{\rm grav}=5 V_0^2 / (\pi G { \rho_0} L_c^2)$, where $V_0$ is the  rms velocity
within the cloud,
representative of the level of turbulent vs gravitational energy in the cloud (eqns.(8-9) of PN).

\subsection{The Hennebelle and Chabrier theory}

In the HC theory of star formation (see HC08, HC09) prestellar cores are the outcome of initial 
density fluctuations that isolate themselves from
the surrounding medium under the action of gravity. These fluctuations are determined by 
identifying in the cloud's random field of density fluctuations the
structures of mass $M$ which at scale $R$ are gravitationally unstable, according 
to the virial theorem. This condition defines a {\it scale-dependent} (log)-density threshold,
$\delta_R^c=\ln(\rho_c(R)/{ \rho_0})$, or equivalently a scale-dependent Jeans mass, $M_R^c$
\begin{eqnarray} 
M_R^c = a_J^{2/3} 
\left( {  (C_s)^2 \over G    } R + {V_0^2  \over 3\, G  } \left({R \over 1 {\rm pc}}\right)^{2\eta} R \right),  
\label{crit_Mtot}
\end{eqnarray}  
where $C_s$ is the sound speed, G the gravitational constant, $a_J$ a constant of order unity 
while $V_0$ and $\eta\simeq0.4$ determine the rms velocity:
\begin{eqnarray}
\langle V_{\rm rms}^2\rangle =  V_0^2 \times \left( {R \over  1 {\rm pc}} \right) ^{2 \eta}.
\label{larson}
 \end{eqnarray} 
 A fluctuation of scale $R$ will be replenished within a typical crossing time
$\tau _R$,  and will thus be 
replenished a number of time equal to $\tau ^0 _{ff} / (\phi_t \tau_{R,ff})$, where
 $\tau_{R,ff}=\tau_R/\phi_t$ is the freefall time at scale $R$   (see appendix),  i.e. at density $\rho_R\sim M_R/R^3$. 
Including this condition into the HC formalism yields, after some algebra, 
for the number-density mass spectrum of gravitationally bound structures, ${\cal N} (M)=d(N/V)/dM$:

\begin{eqnarray}
\label{n_general}
{\cal N} ( M_R)  \simeq
 { {\rho_0} \over M_R} 
{dR \over dM_R} \times \left( -{d \delta_R  \over dR} e^{\delta_R} ({\tau ^0 _{ff} \over \tau_R}) {\cal P}( \delta_R) 
 \right).
\label{spec_mass}
\end{eqnarray}
Apart for the time ratio ${\tau ^0 _{ff} \over \tau_R}$, this equation is similar to 
equation~(33) of HC08 and equation~(27) of HC09.
According to this definition, $SFR_{ff}$ is thus given by the integral of the mass spectrum specified by 
equation~(\ref{spec_mass}):
\begin{eqnarray}
 SFR_{ff} =-
 {\epsilon \over \phi_t} \int _0 ^{M_{cut}} { dM \over M} 
{dR \over dM} \,
{ d \delta_R \over dR} {\tau^0_{ff} \over \tau_{R,ff}} e^{\delta_R} {\cal P}( \delta_R).
\label{sfr_hc}
\end{eqnarray}
According to eqn.~(\ref{crit_Mtot}), $M_{cut}$ correponds to the mass associated with the  largest size fluctuations that can turn unstable in the cloud, 
$y_{cut}=2R/L_c$. We verified that, as long as $y_{cut}$ is not too small, the results depend only weakly on its value (Hennebelle \& Chabrier, in prep.).
In the following, we will pick $y_{cut}\approx 0.1$ as our fiducial value. 

A few remarks are worth discussing at this stage. First, in the present theory there is 
no   explicitly introduced critical scale or density for star formation, as we sum up
over all gravitationally unstable cores, irrespectively 
of their scale or density. This is achieved through the 
multiscale analysis expressed by eqns.~(\ref{spec_mass})-(\ref{sfr_hc}).
Indeed, turbulence is by essence a multi-scale phenomenon and introducing a critical scale
does not appear clearly justified.
Indeed, a piece of fluid, 
even if dominated by turbulence, can still collapse if it is self-gravitating.

Another essential difference with the KM and PN theories  is that these latter rely
 on a unique characteristic collapsing time
(the mean cloud freefall time in KM and the critical 
density freefall time in PN). This can be seen
from  eqns.~(\ref{sfr_simple}) and (\ref{sfr_simple2}), where
the  term $\widetilde{\rho}^{1/2}$, taken at $\widetilde{\rho}_{crit}$,
lies outside the integral. In constrast, in our theory (see eqn.~(\ref{sfr_hc})), the freefall density dependence of each collapsing structure
is properly accounted for, as the freefall time
consistently varies with mass $M$ and scale $R$, $\tau_{R,ff}\propto \rho_R^{-1/2}$.

\subsection{A simplified multi-freefall theory}
\label{sfr_simp}

Even though 
we stress that the SFR cannot be properly determined
by a simple integral of the density PDF,  because such an integral, unlike eqn.~(\ref{sfr_hc}),
does not take into account the spatial distribution of the gas, we suggest the
following simplified but more consistent expression,  which retains the collapsing time
 density-dependence, instead of eqns.~(\ref{sfr_simple}) and ~(\ref{sfr_simple2}):

\begin{eqnarray}
\nonumber
SFR_{ff}^{simp}
 &=& \epsilon
\int ^\infty _{\delta_{crit}} {\tau_{ff}^0 \over \tau_{ff}(\rho) \phi_t} 
 \widetilde{\rho} {\cal P}(\delta) d{\delta} =
{\epsilon  \over  \phi_t}
\int ^\infty _{\delta_{crit}}
\widetilde{\rho}^{3/2} {\cal P}(\delta) d\delta  \\
&=& {\epsilon  \over  2 \phi_t} \exp( 3 \sigma_0^2 / 8 ) 
\left[ 1 + {\rm erf} \left( { \sigma_0^2- \ln( \widetilde{\rho}_{crit} ) \over  
2^{1/2} \sigma_0 }  \right) \right].
\label{sfr_corrected}
\end{eqnarray}
 This SFR is larger than the ones given by eqns.~(\ref{sfr_simple}) and ~(\ref{sfr_simple2}),
 as shown in the next section. 

We consider different choices for $\rho_{crit}$. When using $\rho_{crit,KM}$ or 
 $\rho_{crit,PN}$,
we refer to the corresponding SFR
  as "multi-freefall KM  or PN", respectively.
We also consider another value for $\rho_{crit}$, obtained by simply 
requiring that the Jeans length at this density is  equal to
$y_{cut} L_c$. This corresponds to the assumption that only fluctuations
smaller than a given cloud size fraction can collapse. We simply refer to this 
model as "multi-freefall." 
 It is easy to check that eqn.(\ref{sfr_corrected}) depends only weakly on $y_{cut}$, except when $y_{cut}\rightarrow 0$.

\section{Results}
We now compare 
the  
$SFR_{ff}$ predicted by the various theories and confront the results
to recent observations.

\setlength{\unitlength}{1cm}
\begin{figure*} 
\includegraphics[width=15cm]{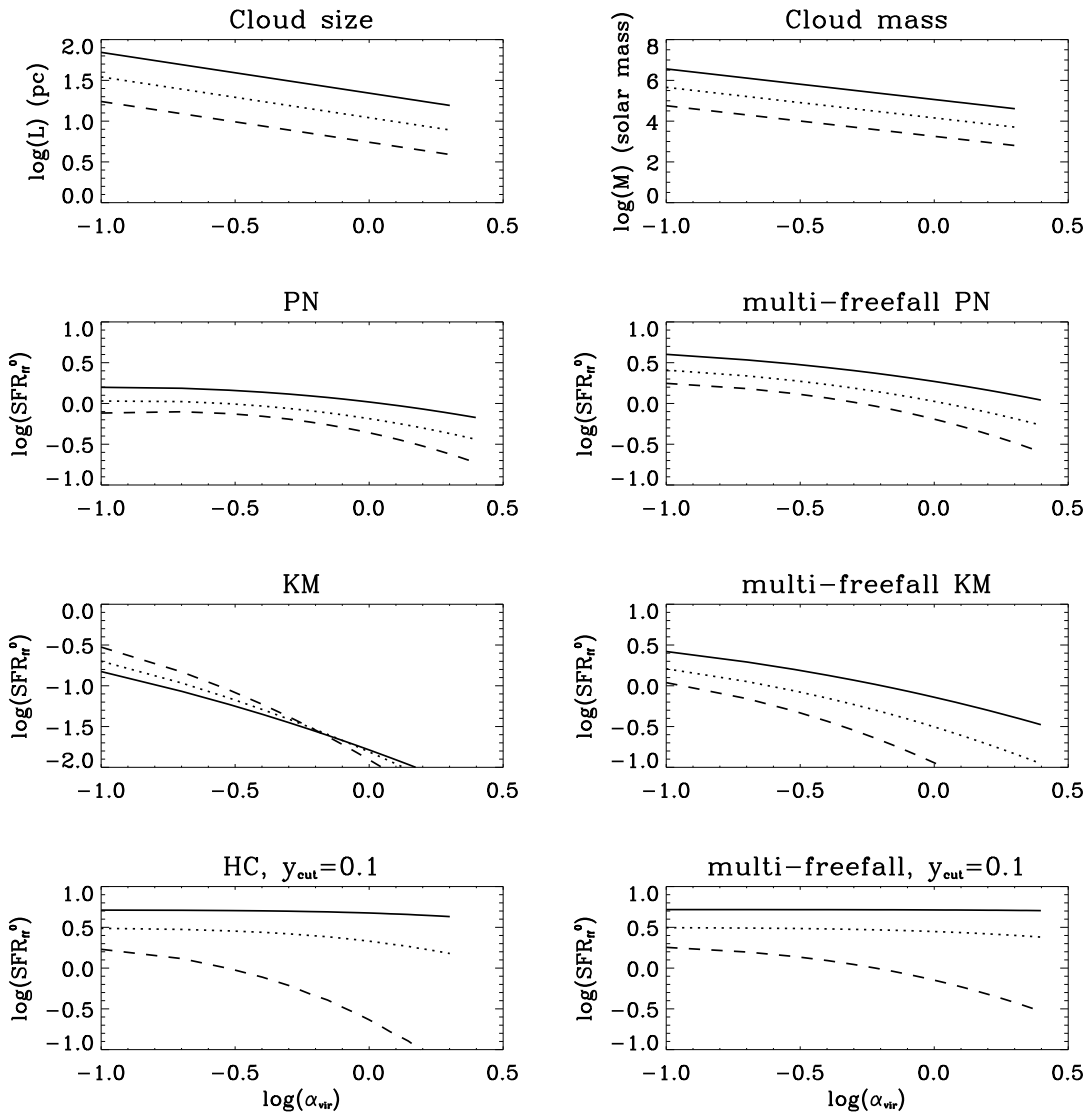}
\caption{$SFR_{ff}^0$ as a function of  $\alpha_{vir}$ for  various
 cloud parameters and ${\cal M}=$ 16 (solid), 9 (dot) and 4 (dash), as predicted by various theories.
Top panels: cloud size and mass.
Second row: Padoan \& Nordlund (2011)(left) and the corresponding "multi-freefall PN" (right).
Third row: Krumholz \& McKee (2005) (left) and "multi-freefall KM" (right); 
Fourth row: Hennebelle \& Chabrier
complete theory (eqn.~(\ref{sfr_hc});left) and simplified "multi-freefall" theory (eqn.~(\ref{sfr_corrected});right) for $y_{cut}=0.1$.}
\label{sfr0}
\end{figure*}

\subsection{Comparison between the various theories}

As in KM and PN, we define the cloud properties by  
$\alpha _{vir}$ and ${\cal M}$.
Figure~\ref{sfr0} displays
 $SFR_{ff}^0$,  which corresponds to $SFR_{ff}$ for $\epsilon=1$ and $\phi_t=1$,
 obtained with different formalisms, 
for various values of the virial parameter, 
and for three typical Mach numbers, namely ${\cal M}=$16, 9 and 4.
 
Both the HC and the multi-freefall models 
 are larger by a factor $\sim$2-3 than PN and by at least an order of magnitude
 than KM. This stems from the fact that, when taking into account the density-dependence of 
the structure collapsing times, (i) dense regions collapse fast, and (ii) fluctuations denser than $\rho_{crit}$ have a smaller 
free-fall time than the ones at $\rho_{crit}$, globally increasing the value of $SFR_{ff}^0$.
When such a density-dependence is properly accounted for, all SFR determinations  are in
better agreement. Nevertheless, some differences persist between
the various multi-freefall models. This stems from the choice of $\rho_{crit}$.
 Indeed, the choice of $\widetilde{\rho}_{crit}$ in the KM and PN theories (see \S2.1 and 2.2) yields 
$y_{cut}\approx {\cal M}^{-2}$ and thus corresponds to
very small values of $y_{cut}$, implying that 
only small Jeans masses (or conversely only very dense structures) are taken into account in these models.

Two interesting trends can be inferred from Fig. 1.
First, increasing the virial parameter leads to a decrease
of the SFR, with a severe reduction above some typical value of $\alpha_{vir}$, which decreases with decreasing Mach number.
This naturally arises from the
fact that, as $\alpha_{vir}$ increases, the increasing contribution of kinetic energy over
 potential energy prevents gravitational collapse and thus inhibits star formation, a point
already noticed by KM and PN. 
Second, the SFR increases, although modestly, with the Mach number.
 This is because increasing the Mach number extends the core mass function (CMF) into the low-mass
 domain (see PN and HC08) and, as 
small-scale structures have shorter free-fall times, this
 increases the number of collapsing small cores, thus the SFR.
This positive dependence of the SFR upon ${\cal M}$ is in agreement with the
results of PN but contrasts with the ones of KM (see their equation~30), 
as seen in the figure. Such a decreasing dependence of the SFR with increasing Mach in the
 KM theory clearly stems from the fact that the $\widetilde{\rho}_{crit}^{1/2}$ term (see eqn.(\ref{sfr_simple2}))
is lacking in their eqn.~(20). 

\subsection{Comparison with observations.}

\begin{figure*} 
\includegraphics[width=15cm]{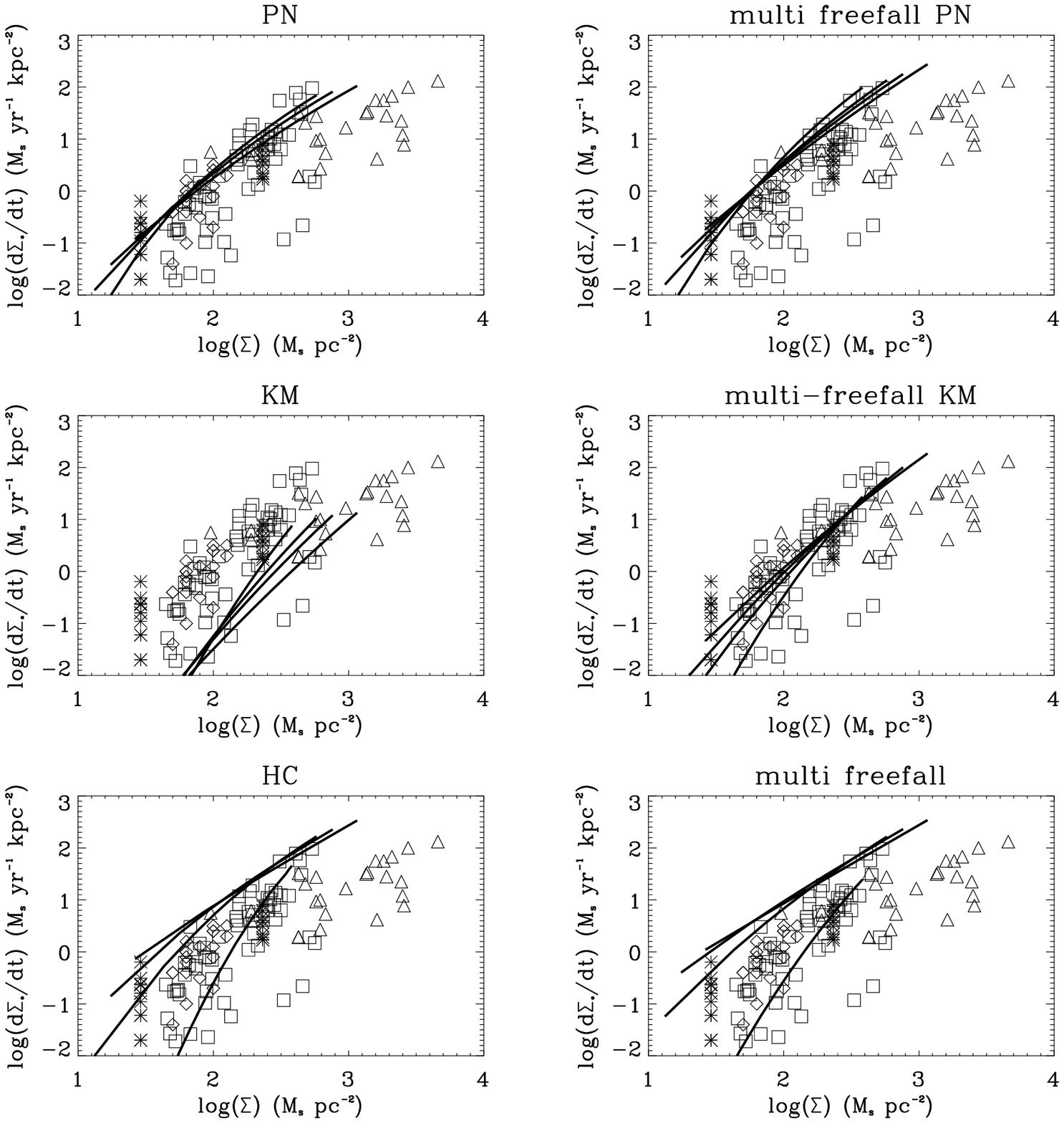}
\caption{Comparison of the SFR per unit area, ${\dot \Sigma}_\star$, as a function of gas
 surface density, ${\Sigma}_g$, as predicted by the various theories illustrated in Fig. 1,
 with the observational determinations of Heiderman et al. (2011) for massive clumps 
(triangles) and molecular clouds (diamonds+squares). The stars show the data of Lada et al. (2010).
The four solid lines correspond to 
four cloud sizes, namely $L_c=1$, 4, 10 and 40 pc (left to right). }
\label{sfr-surf}
\end{figure*}

Star forming
giant molecular clouds in the Milky Way have masses $10^3\la  M_c/M_\odot \la 3 \times 10^6 $, with
 ${\cal M}\approx 4$-30 and $\alpha_{vir} \approx 0.3$-3. The observed SFR per cloud 
free-fall time
lies in the range $0.03\la   SFR_{ff}\la 0.3$, with a mean value 
$\langle SFR_{ff}\rangle\approx 0.16$ (Murray 2011,  although see Feldmann \& Gnedin 2011 for caution). 
Evans et al. (2009) and 
Heiderman et al. (2010) find SFR's in the range $\approx 0.02$-0.12 for
nearby molecular clouds and $\approx 0.03$-0.5 for  massive star-forming dense clumps,
 yielding a mean value $\approx 0.1$,  about an order of magnitude larger than the values predicted by KM.
Krumholz \& Tan (2007, Fig. 5) report lower values at low density. At high density ($\gtrsim 10^{4}$ cm$^{-3}$), however, two of their
three data (ONC and CS(5-4)), are compatible with the aforementioned mean 
values\footnote{It must be kept in mind that all these SFR values apply to giant molecular clouds. 
Values inferred for entire galaxies, including the Milky Way, are substantially lower, as they include diffuse atomic or molecular gas, overestimating the amount of gas counted as star-forming gas (e.g. Heiderman et al. 2010).}.
According to
our calculations (see Fig. 1), for such cloud/clump characteristics, $SFR_{ff}^0$ is
 predicted to lie within the range $\approx 0.3$-3.
As discussed in \S~\ref{krum} and in the appendix, the effective SFR is $SFR_{ff} = (\epsilon / \phi)\times SFR_{ff}^0$, where                   
 $\epsilon/\phi_t\approx $ 0.1-0.2.
 Therefore, according to the present calculations,  theories based on a multi-freefall formalism yield
SFR's per free-fall time in typical molecular clumps 
in the range $SFR_{ff}\approx$ 0.03-0.6, going from low-dense clouds
 to the densest clumps, consistent with the observed values.

Figure 2 displays the SFR per unit area, 
${\dot \Sigma}_\star= SFR_{ff}\times \Sigma_g/\tau^0_{ff}$, with $\Sigma_g = L_c \rho_0=M_c/\pi L_c^2$,
as a function of cloud surface densities, ${\Sigma}_g$, for four typical 
cloud sizes, $L_c=$ 1, 4 10 and 40 pc. The clouds are assumed to follow Larson's (1981) relations 
and thus have velocities given by eqn.~(\ref{larson}) and
densities $n_0 \times (L_c/ 1 {\rm pc})^{-0.7}$, where $n_0=10^2$ to $10^4$ cm$^{-3}$,
 yielding cloud masses $M_c=200$-$10^6$ $M_\odot$. From these values, $\cal M$
and  $\alpha_{vir}$ can be consistently determined.
In Fig. 2, we have taken $\epsilon/\phi_t = 0.1 $.\footnote{This applies to eqn.~(\ref{sfr_simple2}) as well, whereas the true PN relation corresponds to 
 $\epsilon/\phi_t = 1.0 $ and 0.5 in the hydro and MHD case, respectively, and should be moved upward accordingly on the figure.}

 Lada et al. (2010) and Heiderman et al. (2010) data  for clouds and massive clumps are shown 
for comparison. 
Clearly, almost all theories exhibit a direct correlation between the density of star formation 
and the gas density and, when including the $\epsilon/\phi_t $ factor,  reproduce  well the observational results, except possibly 
for the densest clumps, where the SFR's are about a factor $\sim 5$ larger. 
The large spread of the data precludes a clear
distinction  between the theories at this stage,   apart from the KM one which clearly lies well
below most of the data points.

Interestingly,  both the exact HC and "multi free-fall" calculations 
predict a drastic drop in the SFR below 
$\Sigma_c \approx 110$-120 $\msol$ pc$^{-2}$ and a change of slope  in the ${\dot \Sigma}_\star \propto {\Sigma}_g^{N}$ relation, with $N\approx 4.8$ and $N\approx 1.6$, similar to the Kennicutt-Schmidt relation ($N_{KS}\approx 1.4$), respectively below and above $\Sigma_c$, in very good agreement with the 
observations (Heiderman et al. 2010). This density corresponds to a visual extinction
$A_V\approx 6$ ($A_K\approx 1$). A similar
density-threshold for significant dense core population has been identified in 
several surveys (Onishi et al. 1998, Johnstone et al. 2000, Kirk et al. 2006, 
Enoch et al. 2007, Lada et al. 2010, Andr\'e et al. 2010). Various authors 
(e.g. Johnstone et al. 2004, Kirk et al. 2006,
Heiderman et al. 2010) have suggested that the origin of such a density threshold is 
related to magnetic fields, which cannot support the gas against gravitational collapse
 above some density. The present calculations, however, show that there is no need to
 invoke  magnetic field support or MHD shock conditions to get such a threshold although, as mentioned in the appendix,
magnetic fields may contribute {\it dynamically} by reducing the value of $\phi_t$.
The threshold simply stems from the fact that at the corresponding density, 
the size of the clumps becomes comparable to the Jeans length and thus the amount of gas 
appropriate to form stars drops drastically (see HC09 Fig. 8).
The relative similarity between the various "multi free-fall" predictions clearly 
indicates that, besides the $\epsilon / \phi_t$ factor, the key physical quantity which determines the SFR is the 
{\it density-dependence} of the freefall time of the collapsing overdense 
regions induced by turbulence.
An alternative possibility, as recently suggested by Krumholz et al. (2011) is to assume that the 
SFR is simply a constant factor times the cloud's or galaxy's volume density
over mean free-fall time, although a clear physical explanation is lacking at this stage. 

 \section{Conclusion}

We have included the time dependence in our analytical theory of the IMF to determine the SFR. 
The theory, based on a gravoturbulent picture of star formation, yields SFR values in good
 agreement with various observational determinations in Galactic molecular clouds. Moreover, it naturally predicts a 
 density threshold to get significant star formation and yields a dependence of the 
SFR upon gas surface density in very good agreement with the observationally inferred values,  with an abrupt 
change of slope around the threshold. Such a threshold naturally emerges from our theory, without arbitrarily
 introducing a critical density. 

A crucial point at the heart of the present (both complete and simplified (eq. (8))) approach is that,
in contrast to previous theories, the SFR is not characterized by a single, characteristic
 dynamical time in the cloud, but instead involves a {\it density-dependent} collapsing time
 for each turbulence-induced gravity-dominated overdense region in the cloud.
Therefore, in opposite to conclusions based on previous SFR theories, the present results 
show that a SFR determined by turbulence-induced density fluctuations at the early stages of
 star formation provides quite a consistent picture of  star formation in Milky Way 
molecular clouds.  

\section*{Acknowledgement}
GC acknowledges the warm hospitality of the astronomy dpt of the University of Texas, 
where part of this work was conducted, and A. Heiderman and N. Evans for sending their data. 
This research has received funding from the European Research Council under the European
 Community's Seventh Framework Programme (FP7/2007-2013 Grant Agreement no. 247060).

\section*{Appendix: The crossing time}

Our estimate for the crossing time is similar to the estimate 
of Krumholz \& McKee (2005).
The crossing time of a structure of scale $R$ is  $\tau_{ct}(R)=2  R/V_{ct}$. 
At large scales, $V_{ct}\simeq V_{\rm rms}$,
while at small scales, below the sonic length, $V_{ct} \simeq C_s$. 
The typical time $\tau_R$ within which
the density field is significantly modified at scale $R$, 
implying that a new set of fluctuations, statistically 
independent of the former one, has set up is  $\tau_R = \alpha_{ct} \tau_{ct}$,
with $\alpha_{ct}$ a dimensionless coefficient of the order of a few. 
In the Hennebelle-Chabrier theory, we select the pieces of
 gas which are self-gravitating. 
At large scales, this implies $\alpha_g G M / R > V_{\rm rms}^2$,
where $\alpha_g$ is a dimensionless coefficient ($\alpha_g=3/5$ for
a uniform density cloud), while a similar expression holds below the sonic length.  
This yields 
\begin{eqnarray}
\tau_R ={2 \alpha_{ct} R \over V_{\rm rms}  } = 2 \alpha_{ct} 
\sqrt{ 24 \over \pi^2 \alpha_{g}} \tau_{ff}= \phi_t \tau_{ff}=   \phi_t \tau_{ff}^0 \sqrt{{\rho_0}/\rho},\nonumber \\
\label{crossing}
\end{eqnarray}
where $\tau_{ff} $=$\sqrt{3 \pi \over 32 G \rho}$ is the free-fall time of a bound region of density $\rho$, and $\phi_t\approx 3$, yielding $\epsilon/\phi_t\approx0.1$-0.2.

Note that this estimate assumes that it takes about one crossing time to 
rejuvenate a self-gravitating structure. However, it may happen, in particular
in magnetized flows, that all perturbations do not collapse eventually (e.g. Hennebelle \& P\'erault 2000), further increasing $\phi_t$ by a factor of a few.


\end{document}